\documentclass[twocolumn,a4paper,10pt,showpacs,prb]{revtex4-1}
\usepackage[utf8x]{inputenc}
\usepackage{graphicx}
\usepackage{amsmath}
\begin{document}
\title{Superlubric to stick-slip sliding of incommensurate graphene flakes on graphite}
\author{M. M. van Wijk}
\affiliation {Radboud University Nijmegen, Institute for Molecules and Materials, Heyendaalseweg 135, 6525 AJ Nijmegen, The Netherlands}
%\affiliation{Radboud University Nijmegen, P.O. Box 9010, postvak 51, 6500 GL Nijmegen, The Netherlands}
\author{M. Dienwiebel}
\affiliation{Fraunhofer-Institute for Mechanics of Materials IWM, Wöhlerstrasse 11, 79108 Freiburg, Germany}
\affiliation{Institute of Applied Materials (IAM), Karlsruhe Institute of Technology (KIT), Kaiserstraße 12, D-76131 Karlsruhe, Germany}
\author{J. W. M. Frenken}
\affiliation{Kamerlingh Onnes Laboratory, Leiden University, PO Box 9504, 2300 RA Leiden, The Netherlands}
\author{A. Fasolino}
\affiliation {Radboud University Nijmegen, Institute for Molecules and Materials, Heyendaalseweg 135, 6525 AJ Nijmegen, The Netherlands}

\begin{abstract}
We calculate the friction of fully mobile graphene flakes sliding on graphite. For incommensurately stacked flakes, we find a sudden and reversible increase in friction with load, in agreement with experimental observations. The transition from smooth sliding to stick-slip and the corresponding increase in friction is neither due to rotations to commensurate contact nor to dislocations but to a pinning caused by vertical distortions of edge atoms also when they are saturated by Hydrogen. This behavior should apply to all layered materials with strong in-plane bonding. 
\end{abstract}

\pacs{ 68.35.Af,62.20.Qp,68.37.Ps,81.05.uf}
\maketitle

\section{Introduction}
Since the discovery of graphene~\cite{novoselov2004electric}, there is a growing interest in  materials similar to graphite,  with strongly bonded 2D sheets and weak interplanar coupling~\cite{cousin}. This type of bonding makes these lamellar materials, and graphite in particular, also interesting as solid lubricants. 
Since the first Friction Force Microscope (FFM) measurements by Mate et al.~\cite{mate1987atomic} graphite has been a prototype system for nanotribology, the new research field that aims at understanding the fundamental origin of friction at the atomic scale. To model the friction in graphitic materials, it is important to go beyond the Tomlinson model of a point-like tip~\cite{tomlinson1929molecular,prandtl1928gedankenmodell,holscher1998consequences} and consider the tip as an extended contact. In fact, to account for the order of magnitude of the experimentally measured friction on graphite, it was proposed that a graphene flake was dragged along with the tip so that the friction between a graphene flake and graphite was measured instead of that  between tip and graphite~\cite{morita,sasaki,dienwiebel2004superlubricity}. For such an extended contact, the friction strongly depends on the orientation of the flake with respect to the substrate. When the flake has the same orientation as the substrate, the contact is commensurate with high energy barriers to sliding and thus high friction. This is the most energetically favorable configuration, except near the edges of the substrate\cite{Cruz2013}. For angles in between commensurate situations, the flake is approximately incommensurate and the potential barriers to sliding are averaged out. In this situation, an almost frictionless sliding was observed~\cite{hirano1993superlubricity, dienwiebel2004superlubricity,feng2013superlubric}, a phenomenon often referred to as superlubricity~\cite{shinjo,erdemir,gnecco2008superlubricity,ende2012effect}.

Recently, extremely high speed superlubricity has been observed for micron sized graphene flakes~\cite{yang2013observation}. 
However, for finite flakes, superlubricity is not necessarily a stable state. Further experiments and numerical simulations showed that graphene flakes of the order of 100 carbon atoms very often rotate to the commensurate orientation with a large and irreversible  increase of friction~\cite{filippov2008torque}. This finding was further supported by theoretical work, although stable orbits are predicted to exist~\cite{dewijn2010stability}. 

\begin{figure}[htbp]
\centering
 \includegraphics[width=0.7\linewidth]{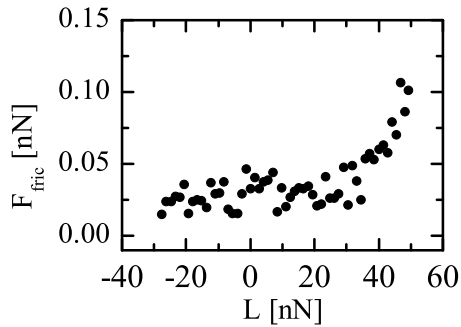}
\caption{FFM measurements of the load dependence of the friction force between a graphene flake on an HOPG graphite surface over a region of the graphite to which the flake is incommensurate. Every data point was the average over 5 separate measurements. The sudden increase of friction at loads in excess of $\sim$ 40 nN  was reproducible and fully reversible.}
 \label{fig:Dienwiebel}
\end{figure}

A sudden increase of friction was also measured for incommensurate graphene flakes with increasing load ~\cite{dienwiebel2003thesis}, as shown in Fig. \ref{fig:Dienwiebel}. Since the observed increase in friction was found to be reversible, it could not be explained by rotations to the commensurate state. Reversibility also rules out plastic deformations of the substrate.  Models with rigid flakes cannot explain this increase since, for incommensurate contacts,  the corrugation remains too flat for the occurrence of stick-slip instabilities~\cite{tomlinson1929molecular,prandtl1928gedankenmodell,gnecco2008superlubricity}. Bonelli et al~\cite{bonelli2009atomistic} did simulations on non-rigid graphene flakes, but they only discussed  the influence of load strongly limiting the deformations of the flake by means of very stiff springs ($K$=2.5 eV/\AA$^2$).

Under load, a breakdown of superlubricity can also occur if strong in-plane distortions lead to local commensurability. This effect was reported by Kim and Falk~\cite{kim2009atomicscale} for a model system of atoms with Lennard-Jones and harmonic interactions. They showed that the tip would adjust to the substrate with higher load or weaker harmonic interaction between tip atoms. This adjustment led to local commensurability and consequently to the breakdown of superlubricity. Here we show that the reversible increase of friction with load shown in Fig. \ref{fig:Dienwiebel} is instead due to pinning of the edge atoms involving mostly vertical motion and very little in-plane strain in the flake. This mechanism seems to be specific of lamellar materials where the creation of defects or dislocations is energetically very unfavorable, contrary to the case of metal and rare gas islands on surfaces where the occurrence of dislocations dominates the diffusion~\cite{hamilton1995dislocation}. 

\section{Model}
\begin{figure}[htbp]
%  \centering
 \includegraphics[width=\linewidth]{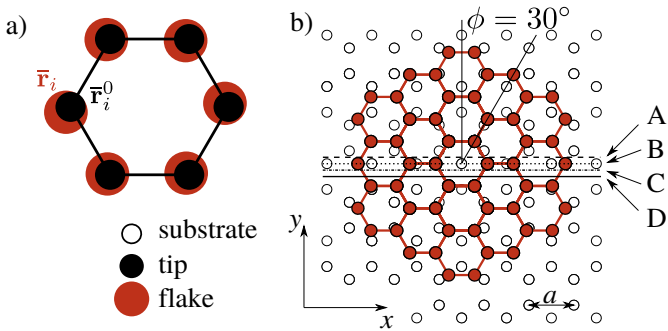}
\caption{(Color online) a) Model of the rigid tip and mobile flake. Each flake atom with coordinate $\mathbf{\bar{r}}_i$ is attached with a spring to a support point $\mathbf{\bar{r}}_i^0$.  b) Flake with $\phi=30 ^{\circ}$ on the substrate. The scanlines over which the center of mass of the support moves, from top to bottom: A, B, C, D. The period $a$ is indicated.}
\label{fig:scanline}
\end{figure}

We construct a model of a FFM experiment where a graphene flake made of $N$ atoms is attached through springs to a tip that is moved on a graphite substrate. We allow the atoms of the flake to move in all directions whereas we keep the substrate atoms at their equilibrium positions in a flat hexagonal lattice at $z=0$. 
The tip is modeled by attaching, in the $x$ and $y$ directions, each atom of the flake to springs  at the positions of a rigid support flake of the same shape as shown in Fig. \ref{fig:scanline}a. Alternatively, one could attach a spring to the center of mass, but this would not limit the rotations of the flake towards a commensurate contact, which is not the physical situation we want to consider. We model the load as a constant force on each atom and report either the load per atom or the total load on the flake given as the sum of the load on all atoms.  
The interaction between the atoms in the flake is given by the REBO potential~\cite{stuart2000reactive,brenner2002second} as implemented in the molecular dynamics (MD) code LAMMPS \cite{lammps}. The equilibrium interatomic distance $d$ is 1.3978~\AA~which is close to the experimental value 1.42~\AA, giving a periodicity in the $x$ direction $a=\sqrt{3}d=2.42$~\AA.
We describe the interlayer interactions in graphitic systems  by means of the  the Kolmogorov-Crespi (KC) potential~\cite{kolmogorov2005registry}. This combination has been shown to accurately reproduce the potential energy surface due to the substrate~\cite{reguz2012potential} which is underestimated by the Lennard-Jones potential in AIREBO~\cite{stuart2000reactive}. With this potential the interplanar distance in graphite is 3.34 \AA, with an energy gain of 48~meV/atom and a difference of 15~meV/atom between AA and AB stacking. 
 
The forces acting on the flake atoms are given by
\begin{align}
 \mathbf{F}_{KC}+\mathbf{F}_{REBO}+\mathbf{F}_{spring}+\mathbf{F}_{load}
\end{align}
where the spring force on atom $i$ is
\begin{align}
\mathbf{F}_{spring, i}=-K(\mathbf{\bar{r}}_i-\mathbf{\bar{r}}_i^0)  
\end{align}
where $\mathbf{\bar{r}}_i$ and $\mathbf{\bar{r}}_i^0$ are the in plane coordinates of the flake atom and its support point respectively, as indicated in Fig. \ref{fig:scanline}a. The spring constant $K$ is taken to be 16~meV/\AA$^2$.

The component in the pulling direction $x$ of the total spring force, 
\begin{align}
F_{x}=\sum_{i=1}^N\mathbf{F}_{spring, i}\cdot \mathbf{\hat{x}}
\end{align} 
is often called the lateral force and its average over time gives the friction force $F_{fric}$. In fact, it can be shown that the average of the lateral force over a period of length $a$ is 
\begin{align}
 <F_{x}>\equiv F_{fric}=\frac{\Delta W}{a}
\end{align}
where $\Delta W$ is equal to the energy dissipated over a period of length $a$ \cite{Annalisaboek}.
In our simulations, we use the first period as a transient and evaluate $F_{fric}$ by averaging over the subsequent three periods. Notice that, in absence of interactions with the substrate, $F_{x}=0$ although each spring force ${F}_{spring, i}$ can be different from zero due to relaxation of the flake induced by the edge termination with respect to the fixed support points. 

The load on atom $i$ is a constant force
\begin{align}
 \mathbf{F}_{load,i}=-L/N \mathbf{\hat{z}}
\end{align}
where $L$ is the total load on the flake. 
 
Since the contact area of the flake has been estimated to be of the order of a hundred atoms~\cite{verhoeven2004model,dienwiebel2004superlubricity}, we choose flakes with hexagonal symmetry consisting of $N=54$, $96$  and $150$ atoms. We will mostly consider the case where the flake is rotated by 30 degrees with respect to the substrate  to ensure incommensurability as shown in Fig.~\ref{fig:scanline}b. This angle corresponds to an incommensurate contact for infinite lattices. By shifting the starting position of the incommensurate flake orthogonally to the pulling direction we examine the different scanlines indicated in Fig.~\ref{fig:scanline}b. 

The speed of a FFM is so low ( $\sim 1-1000$~nm/s) that, to a good approximation, the movement can be considered static. We therefore follow the approach by Bonelli et al.~\cite{bonelli2009atomistic} and use a quasistatic approach in which the support is moved in steps along a given scanline and the flake is relaxed after each step by minimizing the force (Eq.~1) on each atom. We use the FIRE scheme~\cite{bitzek2006fire}, a damped dynamics algorithm, as implemented in LAMMPS~\cite{lammps}. We use 200 minimizations per period, which gives a step size for the movement of the support of 0.012 \AA. The minimization stops when the norm of the global force vector becomes less than 3.12$\cdot10^{-4}$~eV/\AA. We repeat this procedure for 4 periods to confirm periodicity and evaluate $F_{fric}$ as described above. We found that the FIRE scheme is superior to conjugate gradients methods in satisfying the expected periodicity of the motion. 

In comparison with tight binding~\cite{bonelli2009atomistic}, our model is more suitable to study the effect of load due to the longer cut-off. The 3.34~\AA~ interlayer distance in graphite is much longer than the cut-off range used in the tight binding model (2.6~\AA).

The quasistatic approach does not include dynamic and temperature effects. In some cases, we therefore compare results obtained by this method to MD simulations at constant temperature using a Langevin thermostat. We choose a damping parameter $\gamma^{-1}$ of 0.6~ps, a timestep of 1~fs, temperatures of 10~K and 300~K and move the support with a speed of $v=4.84$~m/s.

\section{Results}
\begin{figure}[htbp]
  \centering
 \includegraphics[clip=true,angle=270,width=0.85\linewidth]{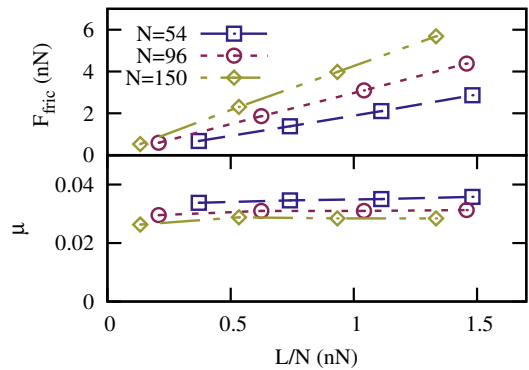}
 \caption{(Color online) Friction and friction coefficient $\mu$ as a function of load for commensurately oriented hexagonal flakes of different sizes as indicated.}
 \label{fig:com}
\end{figure}

In general, incommensurate contacts lead to much lower friction than commensurate ones. In Fig. \ref{fig:com} we show the friction as a function of load for flakes with orientation $\phi=0$, commensurate to the substrate. The friction linearly increases with load, resulting in a nearly constant friction coefficient $\mu=F_{fric}/{L}= 0.03$, which agrees well with the experimental value found for microscale graphene~\cite{marchetto2012friction}. It is worth noting that $\mu$ slightly decreases with size as a result of the fact that atoms at the edges can reach deeper minima and contribute more to the friction. \\
In experiments, the point of zero friction would occur at negative values of load due to the attractive van der Waals forces between tip and substrate. As in our simulations only the contact area of the tip and substrate is modeled, the van der Waals forces are very small and already at $L=0$ the friction is nearly zero.

\begin{figure}[htbp]
 \includegraphics[width=0.9\linewidth,angle=270]{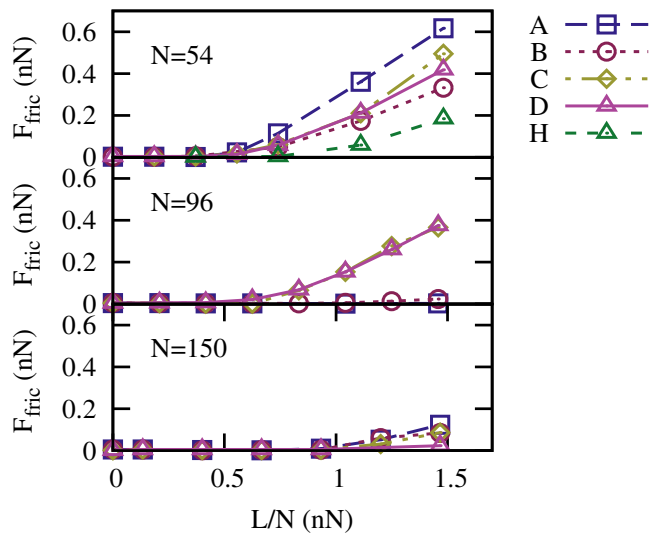}
\caption{(Color online) Friction as a function of load for several scanlines and flake sizes of a flake incommensurately oriented  with $\phi=30 ^{\circ}$ (left). The line labeled 'H' shows the friction of a C$_{54}$H$_{16}$ flake along scanline D.}
\label{fig:Fxvsload}
\end{figure}

\begin{figure}[htbp]
 \includegraphics[clip=true,width=\linewidth]{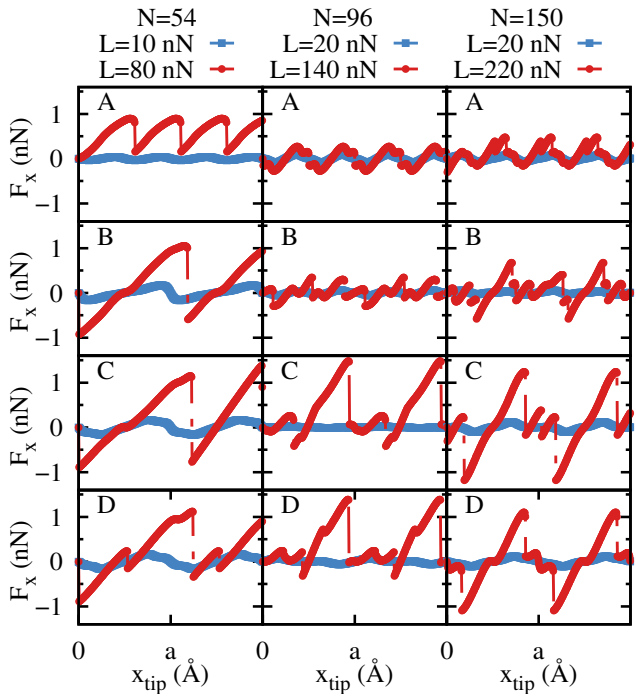}
 \caption{(Color online) Lateral force as a function  of the support position for a low (blue, gray) and a high (red,black) load for the A,B,C,D scanlines (top to bottom) and different sizes (left to right). The total load $L$ is indicated in the figure. Notice the change from smooth to stick-slip behavior.}
 \label{fig:panel}
\end{figure}

The effect of the edges becomes more dramatic for incommensurate contacts. In this case, the friction  is drastically different from the one just discussed for commensurate cases. In Fig. \ref{fig:Fxvsload} we show the variation of the friction force with increasing load for the 54, 96 and 150 atom flakes and the A,B,C,D scanlines. Smaller, 24 atom, flakes are not included because  they rotate to the commensurate orientation already at low  load, due to the smaller moment of inertia. We see that at low load, the friction is indeed almost vanishing as expected for truly incommensurate cases~\cite{peyrard1983critical, ende2012effect}.

At higher loads, not only is the friction at least one order of magnitude lower than for commensurate flakes but the dependence on load is also much more complex. In most cases, the friction suddenly increases from a certain threshold load onwards in a way similar to the experimental situation shown in Fig. \ref{fig:Dienwiebel}a. The increase in friction is stronger, and starts at a slightly lower load, for smaller flakes. For the 96 atom flakes, the increase starts at $L/N\approx0.6$~nN or $L\approx60$~nN, which is close to the experimental value~\cite{dienwiebel2003thesis}. The large increase in friction with load corresponds to the transition from smooth motion to stick-slip behavior shown in Fig. \ref{fig:panel}. For an incommensurate contact, stick-slip motion is very unusual and would only be expected for very small flakes which, strictly speaking, are not incommensurate with the substrate~\cite{clelia}. 

Fig. \ref{fig:Fxvsload} also shows that the behavior with load is not the same for all  scanlines and flake sizes. 
First, we examine in detail the simplest case to explain the behavior with load and the effect of deformations and  thereafter we will discuss the general features for all scanlines and flake sizes. 

\begin{figure}[htbp]
%  \centering
 \includegraphics[width=\linewidth]{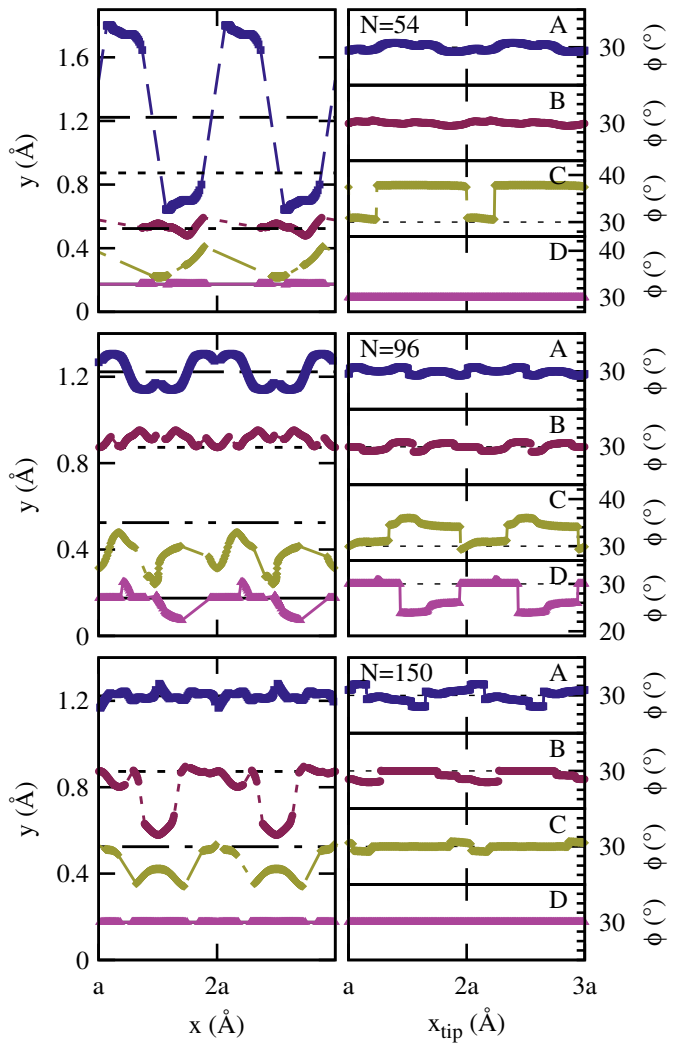}
 \caption{(Color online) Two periods of the trajectory of the center of mass (left) and rotation angle versus the tip position (right). Simulations are for the 54 atom flake with $L=80$ nN or $L/N$=1.48 nN (top), for an 96 atom flake with $L$=140 nN or $L/N$=1.46 nN (middle) and for a 150 atom flake with $L$=220 nN or $L/N$=1.47 nN (bottom). Since the flakes are coupled to the support with springs, the center of mass can deviate from the center of mass of the support (shown in black).}
 \label{fig:trajphi}
\end{figure}

The simplest case is presented by the 54 atom flake along scanline D because the flake remains at 30 degrees and its center of mass does not deviate from the scanline, as shown in Fig. \ref{fig:trajphi}. 
We will focus on the role of deformations of the flake which are usually neglected. 

\begin{figure}[htbp]
 \includegraphics[width=\linewidth]{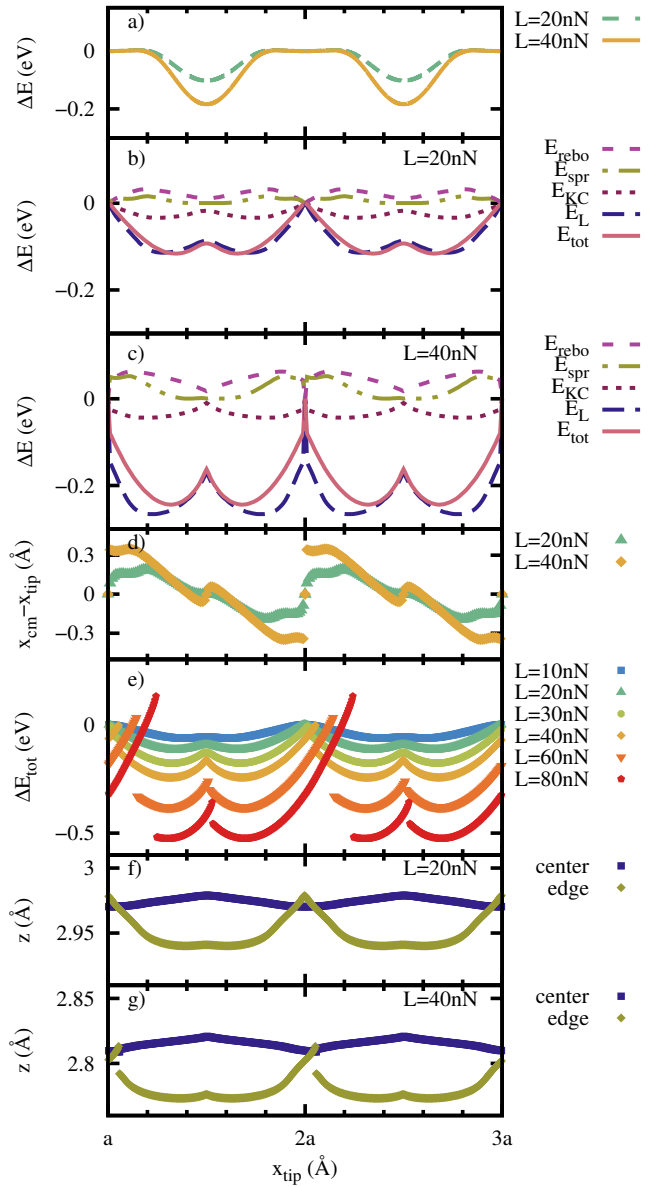}
 \caption{(Color online) 54 atom flake along scanline D: (a)~Energy as a function of the tip position for a rigid flake at a fixed height $z=2.97$~\AA~ and 2.80~\AA~ for $L=20$~nN and $L=40$~nN respectively. (b)~Different contributions to the energy as a function of the tip position for a deformable flake which is relaxed from its ideal configuration for each tip position, for $L= 20$~nN. (c)~Same as in (b) for $L=$40~nN. (d)~Distance between the center of mass of the relaxed flake and the center of mass of the tip. (e)~Variation of the total energy from that at $x_{tip}=0$ for a pulled flake for several loads. (f)~The average distance to the substrate for the edge and non-edge atoms for $L= 20$~nN. (g)~Same as in (f) for $L= 40$~nN.}
 \label{fig:manythingsD54}
\end{figure}

A priori, it is not obvious whether to expect higher or lower friction by considering a non-rigid flake instead of a rigid one. While barriers are easier to circumvent by deforming in addition to shifting and rotating, the same freedom could be used to find deeper minima. 
Therefore we examine the energy profile before looking at the friction. In Fig. \ref{fig:manythingsD54}a we show the potential energy due to the substrate calculated for a rigid flake kept at a  fixed height, corresponding to the average height at the given load. We see that there is a smooth minimum at $a/2$, that becomes deeper with load. To show the importance of the deformation of the flake, we compare this result to the energy profile for a deformable flake. To obtain this, instead of pulling the flake through the coupling to the support, we shift the support with the attached flake and minimize the energy always from the same starting flake configuration.  
In Figs. \ref{fig:manythingsD54}b and \ref{fig:manythingsD54}c it can be seen that the energy profile $E_{tot}$ obtained in this way is very different from the one for a rigid flake (Fig. \ref{fig:manythingsD54}a). This is mainly due to the contribution to the energy associated with load, $E_L=L\cdot z_{CM}$. As load increases from 20~nN (Fig. \ref{fig:manythingsD54}b) to 40~nN (Fig. \ref{fig:manythingsD54}c) the minima in the energy profile of the relaxed flake become deeper and, what is more important, a sharp barrier at $x_{tip}=a$ appears in between at high load.  This barrier gives rise to a discontinuity in the difference between the center of mass of the flake and the support shown in Fig. \ref{fig:manythingsD54}d. When we pull the flake using the quasistatic method described in section II, the sharp barrier at high loads makes the flake stick instead of smoothly following the support. The change of behavior from continuous to stick-slip  is evident in the total energy as a function of tip position in Fig. \ref{fig:manythingsD54}e. At a load of 20~nN the barriers are not curved steeply  enough to pin the flake, resulting in smooth movement, whereas at a load of 40~nN the flake remains pinned in the minima shown in Fig. \ref{fig:manythingsD54}c and the motion becomes discontinuous. As load increases further, a second slip emerges as also the second transition between the two minima is no longer barrierless. The dominant contribution to these energy profiles is the load, as a consequence of atoms being closer or farther away from the substrate. 

\begin{figure}[htbp]
%  \centering
 \includegraphics[width=\linewidth]{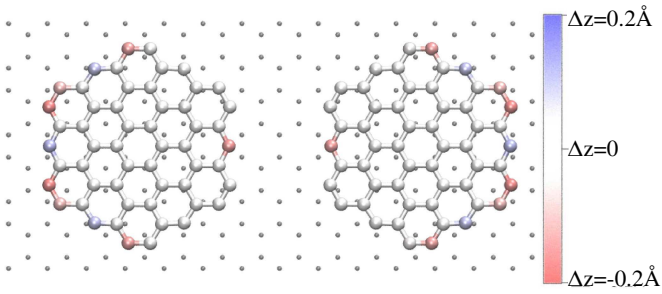}
 \caption{(Color online) Two minimum energy configurations in the trajectory of the 54 atom flake along scanline D for $L= 40$~nN. The color shows $\Delta z=z_i-z_{CM}$ where $_{CM}$ refers to the center of mass of the flake: red is closer to the substrate, blue is farther away. The outer atoms are clearly more mobile.}
 \label{fig:config}
\end{figure}

When we compare the average distance to the surface of the edge atoms, defined as the atoms with two neighbors, to that of the inner atoms (Figs. \ref{fig:manythingsD54}f and \ref{fig:manythingsD54}g), we see that the edge atoms are much closer to the surface and therefore contribute most to the load energy. While the inner atoms move nearly continuously also at higher loads, the edge atoms move  discontinuously. The two configurations corresponding to the two minima in the period $a$ at $L=40$~nN are shown in Fig. \ref{fig:config}. One can see that edge atoms on either side are much closer to the surface and the discontinuity of the motion corresponds in this case to a tilting of the flake in going from $x_{tip}=0.315 a$ to $0.685 a$ and vice versa.

All examined flake sizes and scanlines which show a transition from smooth to stick-slip motion as a function of load, show a corresponding jump in the $z$ coordinates of the edge atoms. As the edge to surface ratio decreases with flake size, edge effects become less important for larger flakes. We indeed observe that the friction increases more, and from a lower load per atom, for smaller flakes, as shown in Fig. \ref{fig:Fxvsload}. 
That the increase in friction at incommensurate orientations is mostly due to edge effects makes friction very dependent on the details of the energy landscape and makes it difficult to draw general trends as a function of scanline, number of particles, and orientation with respect to the substrate. Depending on the scanline, there can be multiple minima or only one per period and the barriers between them can be high or low. For instance for the 96 atom flake, a marked increase of friction is observed for scanlines C and D whereas no or only a very small increase is found for scanlines A and B respectively. The increase also depends on the orientation of the flake. For less symmetric situations, like flakes with $\phi=15 ^{\circ}$ and $\phi=25 ^{\circ}$, we do observe an increase in friction with load for all scanlines. 

\begin{figure}[htbp]
%  \centering
 \includegraphics[width=\linewidth]{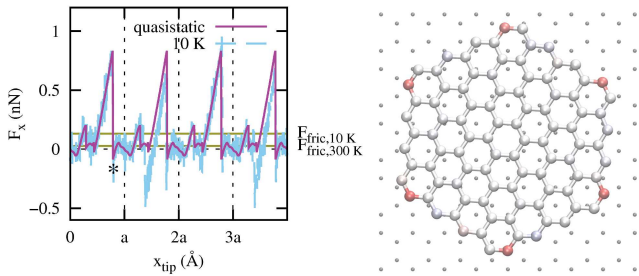}
 \caption{(Color online) Lateral force $F_x$ as a function of the tip position for a 96 atom flake along scanline C for $L=100$~nN or $L/N\approx$ 1~nN calculated by MD at 10~K. The constant lines give the resulting friction at 10~K and the one at 300~K ($F_x$ not shown). The asterisk corresponds to the minimum energy configuration shown in the right figure. The red (dark) color indicates atoms closer to the substrate as in Fig. \ref{fig:config} with  minimum at $\Delta z=-0.165$ \AA.}
 \label{fig:Langevin}
\end{figure}

For the 96 atom flake along scanline C,  we have also performed MD  simulations. In Fig. \ref{fig:Langevin} we see that  at 10~K the flake displays the same stick-slip behavior found in the quasistatic approach. At room temperature however, the stick-slip motion is masked by large fluctuations of the lateral force, resulting in negligible friction. As the speed of nearly 5 m/s is several orders of magnitude higher than in experiments, these simulations are likely to underestimate the friction. The motion of the flake in the stick-slip regime is related to a very symmetric configuration of the flake with the six corner atoms locking into favorable positions of the substrate potential as shown in the right panel of Fig. \ref{fig:Langevin}. 

In an experimental situation, it is likely that the undercoordinated atoms at the edge are saturated by Hydrogen.  To  investigate this situation, we have added Hydrogen atoms to the two-fold bonded edge atoms of the 54 atom flake. We model the H atoms as interacting with the flake and substrate atoms through the REBO potential and neglect interaction with the tip.   The effect of Hydrogen on friction as a function of load for this flake along scanline D is shown by the line labeled 'H' in figure 4. The behavior of friction is qualitatively the same but the increase   starts at a higher load when Hydrogen is present. This fact might have been expected because the in-plane bonds are better preserved up to the edges, making the edges of the flake less flexible.

In addition to the effects studied here, elastic deformations of the substrate could also add to the total friction. The latter effect decreases with the number of layers of the substrate\cite{Carpick, Tewary}. As we model a bulk graphite substrate, we have not included this in our model. However, elastic deformations could significantly contribute to the friction on few-layer graphene substrates.

\section{Summary and conclusion}
In summary, we have studied the motion and friction of mobile and flexible graphene flakes moving at incommensurate orientations on a graphite substrate. In agreement with FFM experimental results, we have found that the superlubric behavior at low loads evolves to a frictional stick-slip motion at high loads. This change is reversible since it is not due to rotations to commensurate orientations but rather to a locking of the flakes as a result of vertical motion of the edge atoms, also when H-saturation is considered.  Interestingly no dislocations appear in the flake also under high load, contrary to the typical behavior of diffusion for metals and rare gas islands on surfaces. The strong in-plane bonding of layered materials like graphite makes that the crystalline structure is preserved while vertical distortions at the edges are energetically favorable. This feature might explain the good and persisting lubricant properties of layered materials. 

\section{Acknowledgements}
We thank Astrid de Wijn, Ted Janssen and Nicola Manini for useful discussions. This work is part of the research program of the Foundation for Fundamental Research on Matter (FOM), which is part of the Netherlands Organisation for Scientific Research (NWO).


\begin{thebibliography}{33}
\expandafter\ifx\csname natexlab\endcsname\relax\def\natexlab#1{#1}\fi
\expandafter\ifx\csname bibnamefont\endcsname\relax
  \def\bibnamefont#1{#1}\fi
\expandafter\ifx\csname bibfnamefont\endcsname\relax
  \def\bibfnamefont#1{#1}\fi
\expandafter\ifx\csname citenamefont\endcsname\relax
  \def\citenamefont#1{#1}\fi
\expandafter\ifx\csname url\endcsname\relax
  \def\url#1{\texttt{#1}}\fi
\expandafter\ifx\csname urlprefix\endcsname\relax\def\urlprefix{URL }\fi
\providecommand{\bibinfo}[2]{#2}
\providecommand{\eprint}[2][]{\url{#2}}

\bibitem[{\citenamefont{Novoselov et~al.}(2004)\citenamefont{Novoselov, Geim,
  Morozov, Jiang, Zhang, Dubonos, Grigorieva, and
  Firsov}}]{novoselov2004electric}
\bibinfo{author}{\bibfnamefont{K.~S.} \bibnamefont{Novoselov}},
  \bibinfo{author}{\bibfnamefont{A.~K.} \bibnamefont{Geim}},
  \bibinfo{author}{\bibfnamefont{S.}~\bibnamefont{Morozov}},
  \bibinfo{author}{\bibfnamefont{D.}~\bibnamefont{Jiang}},
  \bibinfo{author}{\bibfnamefont{Y.}~\bibnamefont{Zhang}},
  \bibinfo{author}{\bibfnamefont{S.}~\bibnamefont{Dubonos}},
  \bibinfo{author}{\bibfnamefont{I.}~\bibnamefont{Grigorieva}},
  \bibnamefont{and} \bibinfo{author}{\bibfnamefont{A.}~\bibnamefont{Firsov}},
  \bibinfo{journal}{Science} \textbf{\bibinfo{volume}{306}},
  \bibinfo{pages}{666} (\bibinfo{year}{2004}).

\bibitem[{\citenamefont{Leb{\`e}gue et~al.}(2013)\citenamefont{Leb{\`e}gue,
  Bj{\"o}rkman, Klintenberg, Nieminen, and Eriksson}}]{cousin}
\bibinfo{author}{\bibfnamefont{S.}~\bibnamefont{Leb{\`e}gue}},
  \bibinfo{author}{\bibfnamefont{T.}~\bibnamefont{Bj{\"o}rkman}},
  \bibinfo{author}{\bibfnamefont{M.}~\bibnamefont{Klintenberg}},
  \bibinfo{author}{\bibfnamefont{R.~M.}~\bibnamefont{Nieminen}}, \bibnamefont{and}
  \bibinfo{author}{\bibfnamefont{O.}~\bibnamefont{Eriksson}},
  \bibinfo{journal}{Phys. Rev. X} \textbf{\bibinfo{volume}{3}},
  \bibinfo{pages}{031002} (\bibinfo{year}{2013}).

\bibitem[{\citenamefont{Mate et~al.}(1987)\citenamefont{Mate, McClelland,
  Erlandsson, and Chiang}}]{mate1987atomic}
\bibinfo{author}{\bibfnamefont{C.~M.} \bibnamefont{Mate}},
  \bibinfo{author}{\bibfnamefont{G.~M.} \bibnamefont{McClelland}},
  \bibinfo{author}{\bibfnamefont{R.}~\bibnamefont{Erlandsson}},
  \bibnamefont{and} \bibinfo{author}{\bibfnamefont{S.}~\bibnamefont{Chiang}},
  \bibinfo{journal}{Phys. Rev. Lett.} \textbf{\bibinfo{volume}{59}},
  \bibinfo{pages}{1942} (\bibinfo{year}{1987}).

\bibitem[{\citenamefont{Tomlinson}(1929)}]{tomlinson1929molecular}
\bibinfo{author}{\bibfnamefont{G.}~\bibnamefont{Tomlinson}},
  \bibinfo{journal}{Phil. Mag.} \textbf{\bibinfo{volume}{7}},
  \bibinfo{pages}{905} (\bibinfo{year}{1929}).

\bibitem[{\citenamefont{Prandtl}(1928)}]{prandtl1928gedankenmodell}
\bibinfo{author}{\bibfnamefont{L.}~\bibnamefont{Prandtl}}, \bibinfo{journal}{Z.
  Angew. Math. Mech} \textbf{\bibinfo{volume}{8}}, \bibinfo{pages}{85}
  (\bibinfo{year}{1928}).

\bibitem[{\citenamefont{H{\"o}lscher et~al.}(1998)\citenamefont{H{\"o}lscher,
  Schwarz, Zw{\"o}rner, and Wiesendanger}}]{holscher1998consequences}
\bibinfo{author}{\bibfnamefont{H.}~\bibnamefont{H{\"o}lscher}},
  \bibinfo{author}{\bibfnamefont{U.~D.}~\bibnamefont{Schwarz}},
  \bibinfo{author}{\bibfnamefont{O.}~\bibnamefont{Zw{\"o}rner}},
  \bibnamefont{and}
  \bibinfo{author}{\bibfnamefont{R.}~\bibnamefont{Wiesendanger}},
  \bibinfo{journal}{Phys. Rev. B} \textbf{\bibinfo{volume}{57}},
  \bibinfo{pages}{2477} (\bibinfo{year}{1998}).

\bibitem[{\citenamefont{Morita et~al.}(1996)\citenamefont{Morita, Fujisawa, and
  Sugawara}}]{morita}
\bibinfo{author}{\bibfnamefont{S.}~\bibnamefont{Morita}},
  \bibinfo{author}{\bibfnamefont{S.}~\bibnamefont{Fujisawa}}, \bibnamefont{and}
  \bibinfo{author}{\bibfnamefont{Y.}~\bibnamefont{Sugawara}},
  \bibinfo{journal}{Surf. Sci. Rep.} \textbf{\bibinfo{volume}{23}},
  \bibinfo{pages}{1} (\bibinfo{year}{1996}).

\bibitem[{\citenamefont{Sasaki et~al.}(1998)\citenamefont{Sasaki, Tsukada,
  Fujisawa, Sugawara, Morita, and Kobayashi}}]{sasaki}
\bibinfo{author}{\bibfnamefont{N.}~\bibnamefont{Sasaki}},
  \bibinfo{author}{\bibfnamefont{M.}~\bibnamefont{Tsukada}},
  \bibinfo{author}{\bibfnamefont{S.}~\bibnamefont{Fujisawa}},
  \bibinfo{author}{\bibfnamefont{Y.}~\bibnamefont{Sugawara}},
  \bibinfo{author}{\bibfnamefont{S.}~\bibnamefont{Morita}}, \bibnamefont{and}
  \bibinfo{author}{\bibfnamefont{K.}~\bibnamefont{Kobayashi}},
  \bibinfo{journal}{Phys. Rev. B} \textbf{\bibinfo{volume}{57}},
  \bibinfo{pages}{3785} (\bibinfo{year}{1998}).

\bibitem[{\citenamefont{Dienwiebel et~al.}(2004)\citenamefont{Dienwiebel,
  Verhoeven, Pradeep, Frenken, Heimberg, and
  Zandbergen}}]{dienwiebel2004superlubricity}
\bibinfo{author}{\bibfnamefont{M.}~\bibnamefont{Dienwiebel}},
  \bibinfo{author}{\bibfnamefont{G.~S.}~\bibnamefont{Verhoeven}},
  \bibinfo{author}{\bibfnamefont{N.}~\bibnamefont{Pradeep}},
  \bibinfo{author}{\bibfnamefont{J.~W.~M.}~\bibnamefont{Frenken}},
  \bibinfo{author}{\bibfnamefont{J.~A.}~\bibnamefont{Heimberg}}, \bibnamefont{and}
  \bibinfo{author}{\bibfnamefont{H.~W.}~\bibnamefont{Zandbergen}},
  \bibinfo{journal}{Phys. Rev. Lett.} \textbf{\bibinfo{volume}{92}},
  \bibinfo{pages}{126101} (\bibinfo{year}{2004}).

\bibitem[{\citenamefont{Cruz-Silva et~al.}(2004)\citenamefont{Cruz-Silva,
  Jia, Terrones, Sumper, Terrones, Dresselhaus, and Meunier}}]{Cruz2013}
\bibinfo{author}{\bibfnamefont{E.}~\bibnamefont{Cruz-Silva}},
  \bibinfo{author}{\bibfnamefont{X.}~\bibnamefont{Jia}},
  \bibinfo{author}{\bibfnamefont{H.}~\bibnamefont{Terrones}},
  \bibinfo{author}{\bibfnamefont{B.~G.}~\bibnamefont{Sumpter}},
  \bibinfo{author}{\bibfnamefont{M.}~\bibnamefont{Terrones}},
  \bibinfo{author}{\bibfnamefont{M.~S.}~\bibnamefont{Dresselhaus}}, \bibnamefont{and}
  \bibinfo{author}{\bibfnamefont{V.}~\bibnamefont{Meunier}},
  \bibinfo{journal}{ACS nano} \textbf{\bibinfo{volume}{7}},
  \bibinfo{pages}{2834} (\bibinfo{year}{2013}).
\bibitem[{\citenamefont{Hirano and Shinjo}(1993)}]{hirano1993superlubricity}
\bibinfo{author}{\bibfnamefont{M.}~\bibnamefont{Hirano}} \bibnamefont{and}
  \bibinfo{author}{\bibfnamefont{K.}~\bibnamefont{Shinjo}},
  \bibinfo{journal}{Wear} \textbf{\bibinfo{volume}{168}}, \bibinfo{pages}{121}
  (\bibinfo{year}{1993}).

\bibitem[{\citenamefont{Feng et~al.}(2013)\citenamefont{Feng, Kwon, Park, and
  Salmeron}}]{feng2013superlubric}
\bibinfo{author}{\bibfnamefont{X.}~\bibnamefont{Feng}},
  \bibinfo{author}{\bibfnamefont{S.}~\bibnamefont{Kwon}},
  \bibinfo{author}{\bibfnamefont{J.~Y.} \bibnamefont{Park}}, \bibnamefont{and}
  \bibinfo{author}{\bibfnamefont{M.}~\bibnamefont{Salmeron}},
  \bibinfo{journal}{ACS Nano} \textbf{\bibinfo{volume}{7}},
  \bibinfo{pages}{1718} (\bibinfo{year}{2013}).

\bibitem[{\citenamefont{Shinjo and Hirano}(1993)}]{shinjo}
\bibinfo{author}{\bibfnamefont{K.}~\bibnamefont{Shinjo}} \bibnamefont{and}
  \bibinfo{author}{\bibfnamefont{M.}~\bibnamefont{Hirano}},
  \bibinfo{journal}{Surf. Sci.} \textbf{\bibinfo{volume}{283}},
  \bibinfo{pages}{473} (\bibinfo{year}{1993}).

\bibitem[{\citenamefont{Erdemir and Martin}(2007)}]{erdemir}
\bibinfo{author}{\bibfnamefont{A.}~\bibnamefont{Erdemir}} \bibnamefont{and}
  \bibinfo{author}{\bibfnamefont{J.-M.} \bibnamefont{Martin}},
  \emph{\bibinfo{title}{Superlubricity}} (\bibinfo{publisher}{Elsevier},
  \bibinfo{year}{2007}).

\bibitem[{\citenamefont{Gnecco et~al.}(2008)\citenamefont{Gnecco, Maier, and
  Meyer}}]{gnecco2008superlubricity}
\bibinfo{author}{\bibfnamefont{E.}~\bibnamefont{Gnecco}},
  \bibinfo{author}{\bibfnamefont{S.}~\bibnamefont{Maier}}, \bibnamefont{and}
  \bibinfo{author}{\bibfnamefont{E.}~\bibnamefont{Meyer}}, \bibinfo{journal}{J.
  Phys.: Condens. Matter} \textbf{\bibinfo{volume}{20}},
  \bibinfo{pages}{354004} (\bibinfo{year}{2008}).

\bibitem[{\citenamefont{van~den Ende et~al.}(2012)\citenamefont{van~den Ende,
  de~Wijn, and Fasolino}}]{ende2012effect}
\bibinfo{author}{\bibfnamefont{J.~A.} \bibnamefont{van~den Ende}},
  \bibinfo{author}{\bibfnamefont{A.~S.} \bibnamefont{de~Wijn}},
  \bibnamefont{and} \bibinfo{author}{\bibfnamefont{A.}~\bibnamefont{Fasolino}},
  \bibinfo{journal}{J. Phys.: Condens. Matter} \textbf{\bibinfo{volume}{24}},
  \bibinfo{pages}{445009} (\bibinfo{year}{2012}).

\bibitem[{\citenamefont{Yang et~al.}(2013)\citenamefont{Yang, Liu, Grey, Xu,
  Li, Liu, Urbakh, Cheng, and Zheng}}]{yang2013observation}
\bibinfo{author}{\bibfnamefont{J.}~\bibnamefont{Yang}},
  \bibinfo{author}{\bibfnamefont{Z.}~\bibnamefont{Liu}},
  \bibinfo{author}{\bibfnamefont{F.}~\bibnamefont{Grey}},
  \bibinfo{author}{\bibfnamefont{Z.}~\bibnamefont{Xu}},
  \bibinfo{author}{\bibfnamefont{X.}~\bibnamefont{Li}},
  \bibinfo{author}{\bibfnamefont{Y.}~\bibnamefont{Liu}},
  \bibinfo{author}{\bibfnamefont{M.}~\bibnamefont{Urbakh}},
  \bibinfo{author}{\bibfnamefont{Y.}~\bibnamefont{Cheng}}, \bibnamefont{and}
  \bibinfo{author}{\bibfnamefont{Q.}~\bibnamefont{Zheng}},
  \bibinfo{journal}{Phys. Rev. Lett.} \textbf{\bibinfo{volume}{110}},
  \bibinfo{pages}{255504} (\bibinfo{year}{2013}).

\bibitem[{\citenamefont{Filippov et~al.}(2008)\citenamefont{Filippov,
  Dienwiebel, Frenken, Klafter, and Urbakh}}]{filippov2008torque}
\bibinfo{author}{\bibfnamefont{A.~E.}~\bibnamefont{Filippov}},
  \bibinfo{author}{\bibfnamefont{M.}~\bibnamefont{Dienwiebel}},
  \bibinfo{author}{\bibfnamefont{J.~W.~M.}~\bibnamefont{Frenken}},
  \bibinfo{author}{\bibfnamefont{J.}~\bibnamefont{Klafter}}, \bibnamefont{and}
  \bibinfo{author}{\bibfnamefont{M.}~\bibnamefont{Urbakh}},
  \bibinfo{journal}{Phys. Rev. Lett.} \textbf{\bibinfo{volume}{100}},
  \bibinfo{pages}{046102} (\bibinfo{year}{2008}).

\bibitem[{\citenamefont{de~Wijn et~al.}(2010)\citenamefont{de~Wijn, Fusco, and
  Fasolino}}]{dewijn2010stability}
\bibinfo{author}{\bibfnamefont{A.~S.}~\bibnamefont{de~Wijn}},
  \bibinfo{author}{\bibfnamefont{C.}~\bibnamefont{Fusco}}, \bibnamefont{and}
  \bibinfo{author}{\bibfnamefont{A.}~\bibnamefont{Fasolino}},
  \bibinfo{journal}{Phys. Rev. E} \textbf{\bibinfo{volume}{81}},
  \bibinfo{pages}{046105} (\bibinfo{year}{2010}).

\bibitem[{\citenamefont{Dienwiebel}(2003)}]{dienwiebel2003thesis}
\bibinfo{author}{\bibfnamefont{M.}~\bibnamefont{Dienwiebel}}, Ph.D. thesis,
  \bibinfo{school}{Leiden University} (\bibinfo{year}{2003}).

\bibitem[{\citenamefont{Bonelli et~al.}(2009)\citenamefont{Bonelli, Manini,
  Cadelano, and Colombo}}]{bonelli2009atomistic}
\bibinfo{author}{\bibfnamefont{F.}~\bibnamefont{Bonelli}},
  \bibinfo{author}{\bibfnamefont{N.}~\bibnamefont{Manini}},
  \bibinfo{author}{\bibfnamefont{E.}~\bibnamefont{Cadelano}}, \bibnamefont{and}
  \bibinfo{author}{\bibfnamefont{L.}~\bibnamefont{Colombo}},
  \bibinfo{journal}{Eur. Phys. J. B} \textbf{\bibinfo{volume}{70}},
  \bibinfo{pages}{449} (\bibinfo{year}{2009}).

\bibitem[{\citenamefont{Kim and Falk}(2009)}]{kim2009atomicscale}
\bibinfo{author}{\bibfnamefont{W.~K.}~\bibnamefont{Kim}} \bibnamefont{and}
  \bibinfo{author}{\bibfnamefont{M.~L.}~\bibnamefont{Falk}},
  \bibinfo{journal}{Phys. Rev. B} \textbf{\bibinfo{volume}{80}},
  \bibinfo{pages}{235428} (\bibinfo{year}{2009}).

\bibitem[{\citenamefont{Hamilton et~al.}(1995)\citenamefont{Hamilton, Daw, and
  Foiles}}]{hamilton1995dislocation}
\bibinfo{author}{\bibfnamefont{J.~C.}~\bibnamefont{Hamilton}},
  \bibinfo{author}{\bibfnamefont{M.~S.}~\bibnamefont{Daw}}, \bibnamefont{and}
  \bibinfo{author}{\bibfnamefont{S.~M.}~\bibnamefont{Foiles}},
  \bibinfo{journal}{Phys. Rev. Lett.} \textbf{\bibinfo{volume}{74}},
  \bibinfo{pages}{2760} (\bibinfo{year}{1995}).

\bibitem[{\citenamefont{Stuart et~al.}(2000)\citenamefont{Stuart, Tutein, and
  Harrison}}]{stuart2000reactive}
\bibinfo{author}{\bibfnamefont{S.}~\bibnamefont{Stuart}},
  \bibinfo{author}{\bibfnamefont{A.}~\bibnamefont{Tutein}}, \bibnamefont{and}
  \bibinfo{author}{\bibfnamefont{J.}~\bibnamefont{Harrison}},
  \bibinfo{journal}{J. Chem. Phys.} \textbf{\bibinfo{volume}{112}},
  \bibinfo{pages}{6472} (\bibinfo{year}{2000}).

\bibitem[{\citenamefont{Brenner et~al.}(2002)\citenamefont{Brenner, Shenderova,
  Harrison, Stuart, Ni, and Sinnott}}]{brenner2002second}
\bibinfo{author}{\bibfnamefont{D.}~\bibnamefont{Brenner}},
  \bibinfo{author}{\bibfnamefont{O.}~\bibnamefont{Shenderova}},
  \bibinfo{author}{\bibfnamefont{J.}~\bibnamefont{Harrison}},
  \bibinfo{author}{\bibfnamefont{S.}~\bibnamefont{Stuart}},
  \bibinfo{author}{\bibfnamefont{B.}~\bibnamefont{Ni}}, \bibnamefont{and}
  \bibinfo{author}{\bibfnamefont{S.}~\bibnamefont{Sinnott}},
  \bibinfo{journal}{J. Phys.: Condens. Matter} \textbf{\bibinfo{volume}{14}},
  \bibinfo{pages}{783} (\bibinfo{year}{2002}).

\bibitem[{\citenamefont{Plimpton et~al.}(1995)}]{lammps}
\bibinfo{author}{\bibfnamefont{S.}~\bibnamefont{Plimpton}}, 
\bibinfo{journal}{J. Comput. Phys.}
  \textbf{\bibinfo{volume}{117}}, \bibinfo{pages}{1} (\bibinfo{year}{1995}).

\bibitem[{\citenamefont{Kolmogorov and Crespi}(2005)}]{kolmogorov2005registry}
\bibinfo{author}{\bibfnamefont{A.~N.}~\bibnamefont{Kolmogorov}} \bibnamefont{and}
  \bibinfo{author}{\bibfnamefont{V.~H.}~\bibnamefont{Crespi}},
  \bibinfo{journal}{Phys. Rev. B} \textbf{\bibinfo{volume}{71}},
  \bibinfo{pages}{235415} (\bibinfo{year}{2005}).

\bibitem[{\citenamefont{Reguzzoni et~al.}(2012)\citenamefont{Reguzzoni,
  Fasolino, Molinari, and Righi}}]{reguz2012potential}
\bibinfo{author}{\bibfnamefont{M.}~\bibnamefont{Reguzzoni}},
  \bibinfo{author}{\bibfnamefont{A.}~\bibnamefont{Fasolino}},
  \bibinfo{author}{\bibfnamefont{E.}~\bibnamefont{Molinari}}, \bibnamefont{and}
  \bibinfo{author}{\bibfnamefont{M.~C.}~\bibnamefont{Righi}},
  \bibinfo{journal}{Phys. Rev. B} \textbf{\bibinfo{volume}{86}},
  \bibinfo{pages}{245434} (\bibinfo{year}{2012}).

\bibitem[{\citenamefont{Fasolino}()}]{Annalisaboek}
\bibinfo{author}{\bibfnamefont{A.}~\bibnamefont{Fasolino}}, in
  \emph{\bibinfo{title}{Handbook of theoretical and computational
  nanotechnology}}, edited by M. Rieth and W. Schommers (\bibinfo{publisher}{American Scientific Publishers}, \bibinfo{year}{2006}).

\bibitem[{\citenamefont{Verhoeven et~al.}(2004)\citenamefont{Verhoeven,
  Dienwiebel, and Frenken}}]{verhoeven2004model}
\bibinfo{author}{\bibfnamefont{G.~S.} \bibnamefont{Verhoeven}},
  \bibinfo{author}{\bibfnamefont{M.}~\bibnamefont{Dienwiebel}},
  \bibnamefont{and} \bibinfo{author}{\bibfnamefont{J.~W.~M.}
  \bibnamefont{Frenken}}, \bibinfo{journal}{Phys. Rev. B}
  \textbf{\bibinfo{volume}{70}}, \bibinfo{pages}{165418}
  (\bibinfo{year}{2004}).

\bibitem[{\citenamefont{Bitzek et~al.}(2006)\citenamefont{Bitzek, Koskinen,
  G{\"a}hler, Moseler, and Gumbsch}}]{bitzek2006fire}
\bibinfo{author}{\bibfnamefont{E.}~\bibnamefont{Bitzek}},
  \bibinfo{author}{\bibfnamefont{P.}~\bibnamefont{Koskinen}},
  \bibinfo{author}{\bibfnamefont{F.}~\bibnamefont{G{\"a}hler}},
  \bibinfo{author}{\bibfnamefont{M.}~\bibnamefont{Moseler}}, \bibnamefont{and}
  \bibinfo{author}{\bibfnamefont{P.}~\bibnamefont{Gumbsch}},
  \bibinfo{journal}{Phys. Rev. Lett.} \textbf{\bibinfo{volume}{97}},
  \bibinfo{pages}{170201} (\bibinfo{year}{2006}).

\bibitem[{\citenamefont{Marchetto et~al.}(2012)\citenamefont{Marchetto, Held,
  Hausen, W{\"a}hlisch, Dienwiebel, and Bennewitz}}]{marchetto2012friction}
\bibinfo{author}{\bibfnamefont{D.}~\bibnamefont{Marchetto}},
  \bibinfo{author}{\bibfnamefont{C.}~\bibnamefont{Held}},
  \bibinfo{author}{\bibfnamefont{F.}~\bibnamefont{Hausen}},
  \bibinfo{author}{\bibfnamefont{F.}~\bibnamefont{W{\"a}hlisch}},
  \bibinfo{author}{\bibfnamefont{M.}~\bibnamefont{Dienwiebel}},
  \bibnamefont{and}
  \bibinfo{author}{\bibfnamefont{R.}~\bibnamefont{Bennewitz}},
  \bibinfo{journal}{Tribol. Lett.} \textbf{\bibinfo{volume}{48}},
  \bibinfo{pages}{77} (\bibinfo{year}{2012}).

\bibitem[{\citenamefont{Peyrard and Aubry}(1983)}]{peyrard1983critical}
\bibinfo{author}{\bibfnamefont{M.}~\bibnamefont{Peyrard}} \bibnamefont{and}
  \bibinfo{author}{\bibfnamefont{S.}~\bibnamefont{Aubry}}, \bibinfo{journal}{J.
  Phys. C} \textbf{\bibinfo{volume}{16}}, \bibinfo{pages}{1593}
  (\bibinfo{year}{1983}).

\bibitem[{\citenamefont{Reguzzoni and Righi}(2012)}]{clelia}
\bibinfo{author}{\bibfnamefont{M.}~\bibnamefont{Reguzzoni}} \bibnamefont{and}
  \bibinfo{author}{\bibfnamefont{M.~C.}~\bibnamefont{Righi}},
  \bibinfo{journal}{Phys. Rev. B} \textbf{\bibinfo{volume}{85}},
  \bibinfo{pages}{201412} (\bibinfo{year}{2012}).

\bibitem[{\citenamefont{Carpick}(2012)}]{Carpick}
\bibinfo{author}{\bibfnamefont{C.}~\bibnamefont{Lee}},
\bibinfo{author}{\bibfnamefont{Q.}~\bibnamefont{Li}},
\bibinfo{author}{\bibfnamefont{W.}~\bibnamefont{Kalb}},
\bibinfo{author}{\bibfnamefont{X.-Z.}~\bibnamefont{Liu}},
\bibinfo{author}{\bibfnamefont{H.}~\bibnamefont{Berger}},
\bibinfo{author}{\bibfnamefont{R.~W.}~\bibnamefont{Carpick}},
 \bibnamefont{and}
  \bibinfo{author}{\bibfnamefont{J.}~\bibnamefont{Hone}},
  \bibinfo{journal}{Science} \textbf{\bibinfo{volume}{328}},
  \bibinfo{pages}{76} (\bibinfo{year}{2010}).

\bibitem[{\citenamefont{}(2012)}]{Tewary}
\bibinfo{author}{\bibfnamefont{A.}~\bibnamefont{Smolyanitsky}}, 
\bibinfo{author}{\bibfnamefont{J.~P.}~\bibnamefont{Killgore}}, 
\bibnamefont{and}
\bibinfo{author}{\bibfnamefont{V.~K.}~\bibnamefont{Tewary}},
  \bibinfo{journal}{Phys. Rev. B} \textbf{\bibinfo{volume}{85}},
  \bibinfo{pages}{035412} (\bibinfo{year}{2012}).


\end{thebibliography}
\end{document}